\documentclass[reprint,aps,prl,superscriptaddress,longbibliography,floatfix]{revtex4-2}
\usepackage{amsmath, amssymb}
\usepackage{graphicx}
\usepackage{dcolumn} 
\usepackage{bm}      
\usepackage{booktabs}
\usepackage{physics}
\usepackage{units}

\newcommand{\ie}{\textit{i.e.,}\ }
\newcommand{\eg}{\textit{e.g.}\ }
\newcommand{\etal}{\textit{et al.}}

\emergencystretch=\maxdimen
\begin{document}

\title{Single-Shot Intensity-Correlation Diffractive X-ray Imaging of ICF Plasmas}

\author{Kenan Qu}
\affiliation{Department of Astrophysical Sciences, Princeton University, Princeton, New Jersey 08544, USA}
\author{Daniel Bhatti}
\affiliation{Networked Quantum Devices Unit, Okinawa Institute of Science and Technology Graduate University, Okinawa 904-0495, Japan}
\author{Nathaniel J. Fisch}
\affiliation{Department of Astrophysical Sciences, Princeton University, Princeton, New Jersey 08544, USA}

\begin{abstract}
X-ray radiography of inertial confinement fusion plasmas is currently limited to several-micron resolution by geometric blur, diffraction, and photon-throughput tradeoffs. We propose single-shot intensity-correlation diffractive imaging (IDI) as a lensless route to submicron plasma turbulence measurements under low-self-emission conditions. Rather than relying on physical apertures, IDI reconstructs plasma morphology by Fourier transforming the spatial correlations of chaotic far-field speckles via the Hanbury Brown--Twiss effect. The Fourier phase is retrieved by applying bispectral closure-phase constraints derived from third-order intensity correlations. We demonstrate this submicron capability in a numerical simulation using a $50~\mathrm{keV}$ x-ray probe scattered by a spiral plasma structure.
\end{abstract}

\maketitle

\textit{Introduction---}%
X-ray radiography is the primary diagnostic used to infer the shape, symmetry, and dynamics of inertial confinement fusion (ICF) implosions~\cite{edwards2013progress, hurricane2014fuel, lindl1995development}. While existing pinhole and point-projection imagers are highly optimized, their spatial resolution is constrained by geometric and diffractive tradeoffs. Resolving smaller features requires a smaller physical aperture, which inevitably starves the detector of photons and exacerbates diffraction blur from the aperture edges~\cite{nugent2010coherent,Kozioziemski2023,koch2012xray,landen2001xray}. At present, the practical resolution limit for hard x-ray ICF radiography is capped at approximately $5~\mu\mathrm{m}$.

Accessing the submicron regime is critical for bridging macroscopic implosion asymmetries and dissipative transport scales. Current diagnostic technique blurs the micron-scale density spikes and vortex sheets into uniform mix layers, obscuring the morphology of, \eg Rayleigh--Taylor or Kelvin--Helmholtz instabilities~\cite{bodner1974rayleigh, kilkenny1994ablative}. These fine-scale instabilities drive turbulent mixing at the fuel-ablator interface, which is considered a major cause of yield degradation~\cite{haan1989onset, atzeni2004physics}. 
It could also contribute to addressing the long-standing controversy about ion opacity at solar-interior conditions~\cite{asplund2009, bailey2015, hoarty2023}. 
Experiments by Heeter \etal~\cite{heeter2017conceptual, heeter2018} rely on the assumption that laser-heated samples expand into perfectly uniform plasmas, but unrecognized small-scale turbulence could skew these results. 
Furthermore, resolving these microscopic eddies is now essential for testing emerging kinetic reactivity models. Recent theories~\cite{Fetsch2025a,Fetsch2025b} indicate that because fast ions sample velocity gradients across adjacent eddies, overall fusion reactivity~\cite{craxton2015direct} is highly sensitive to the exact submicron turbulent flow topology rather than just bulk thermal temperature. By extracting a submicron projected-density spectrum, the diagnostic proposed here could establish plasma turbulence as a measurable state variable in ICF implosions.

The difficulty associated with ICF plasma morphology is not merely the absence of x-ray lenses. Fresnel zone plates and related diffractive optics can theoretically reach smaller length scales, but their high-aspect-ratio fabrication, multiple diffraction orders, and survivability near ICF targets make them poorly matched to ICF radiography~\cite{robinson2001coherent, Seiboth2010hard,abbey2008lensless}. Most importantly, these optics require coherent backlighters.
In the pre- and post-stagnation stages emphasized here, the relative absence of plasma self-emission makes background subtraction a secondary concern compared to this need for coherence. Currently, short-pulse laser-driven hard x-ray backlighters are the optimal choice to provide the picosecond temporal gate needed to freeze hydrodynamic motion and the photon energy needed to penetrate dense targets. Because these sources are inherently spatially incoherent, the central challenge is extracting submicron spatial information without sacrificing vital photon flux to a microscopic image-forming aperture.

Here, we propose overcoming this limitation by adopting single-shot intensity-correlation diffractive imaging (IDI)~\cite{classen2017incoherent, schneider2018x, Trost2023imaging}. Rather than relying on physical x-ray lenses, IDI shifts the spatial-resolution burden entirely onto the detector. By recording the unconstrained far-field speckle pattern across a large-area pixel array the diagnostic allows broad-bandwidth signal collection without a pinhole. Submicron resolution has previously eluded existing backlighters because conventional radiography relies on geometric magnification, which requires restrictively small apertures. In contrast, IDI bypasses this constraint by relying on a high total pixel count to capture spatial correlations over a wide angular acceptance. For example, a $50~\mathrm{keV}$ hard x-ray backlighter  recorded on a $10~\mathrm{cm}$ detector at $3.5~\mathrm{m}$ yields a maximum half-angle corresponding to an Abbe limit of $1.7~\mathrm{nm}$. Although this resolution is reduced in practice due to the statistical signal-to-noise ratios, the effective single-shot resolution still readily achieves submicron scale, demonstrating that the geometric aperture is no longer the fundamental obstacle.

While diffractive imaging typically requires coherent backlighters to form macroscopic interference fringes, IDI bypasses this requirement by using statistical intensity fluctuations similar to the Hanbury Brown--Twiss intensity interferometry~\cite{hbt1956correlation,glauber1963quantum,goodman1976some}. Even when the incident hard x-ray field is chaotic, the image is encoded in the far-field intensity speckle patterns. The second-order spatial covariance of the speckles produces the modulus of the Fourier modes, while third-order correlations provide bispectral closure-phase constraints for the Fourier phases. Crucially, unlike the original HBT measurements that exploit temporal correlations across multiple independent exposures, ICF plasma dynamics are far too rapid, and x-ray detector integration times far too slow, to permit time-domain ensemble averaging. Therefore, IDI relies entirely on spatial ergodicity; in a single shot, the ensemble average is estimated by aggregating statistically independent detector-pixel pairs or triplets with common reciprocal-space separations. The image resolution is no longer tied directly to a physical aperture but to the highest reciprocal-space bandwidth over which the measured correlations are statistically reliable. Thus this approach offers a practical framework for extending single-shot ICF plasma x-ray diagnostics beyond conventional pinhole-resolution limits using existing hard x-ray backlighters.


\textit{SAXS and Diffractive Imaging---}%
Because the hydrodynamic structures of interest are far larger than the hard x-ray wavelength $\lambda$, their scattered signal is concentrated at small angles, placing the measurement in the small-angle x-ray scattering (SAXS) regime~\cite{glenzer2009thomson, glenzer2016matter}. In the weak-phase limit, a probe propagating along $z$ acquires the projected phase shift
\begin{equation}
    \Delta\phi(\mathbf r)=-r_e\lambda\int \delta n_e(\mathbf r,z)\,dz,
    \label{eq:phase}
\end{equation}
where $r_e$ is the classical electron radius, and $\delta n_e$ is the plasma density fluctuation.  The transmission is $T(\mathbf r)=\exp[i\Delta\phi(\mathbf r)] \simeq1+i\Delta\phi(\mathbf r)$ for $|\Delta\phi|\ll1$.
After the transmitted beam $E_{\rm in}$ is blocked by a beam stop, the far-field scattered amplitude is
\begin{equation}
    E_s(\mathbf q) = i\int E_{\rm in}(\mathbf r)\Delta\phi(\mathbf r)
    e^{-i\mathbf q\cdot\mathbf r}\,d^2\mathbf r
    \label{eq:field}
\end{equation}
with $q\simeq 2\pi\theta/\lambda$ for small scattering angle $\theta$. If the incident field is coherent and uniform, Eq.~(\ref{eq:field}) reduces to $E_s(\mathbf q)=iE_{\rm in}\widetilde{\Delta\phi}(\mathbf q)$, where a tilde denotes a Fourier transform. The detected coherent diffraction pattern, $I(\mathbf q)\propto |\widetilde{\Delta\phi}(\mathbf q)|^2$, could then be used for coherent diffractive imaging (CDI) assisted by phase-retrieval algorithms using support and object constraints~\cite{fienup1982phase,miao1999extending}.

\begin{figure}[t]
    \centering
    \includegraphics[width=\columnwidth]{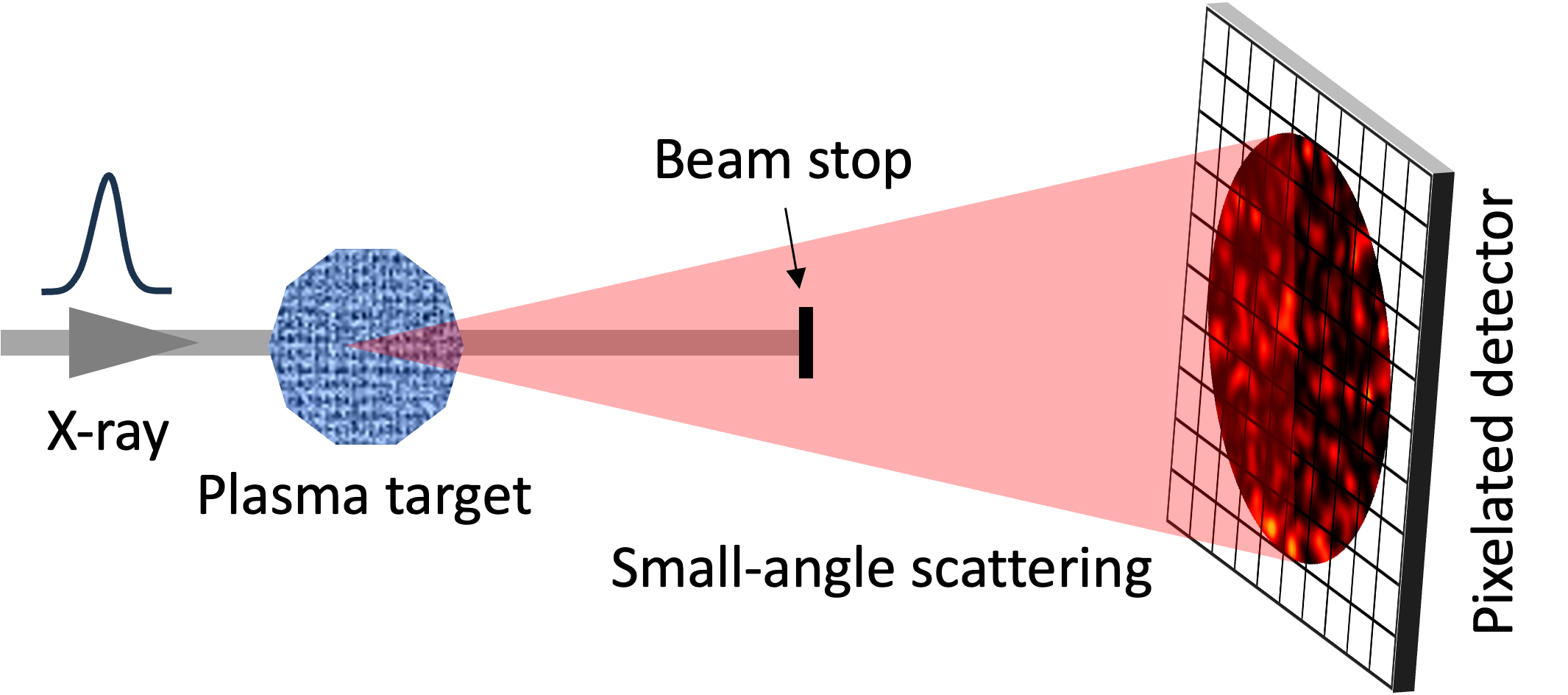} 
    \caption{
    Schematic of the proposed diffractive x-ray imaging geometry.
    }
    \label{fig:setup}
\end{figure}

By sampling object spatial frequencies rather than projecting a magnified ray image, diffractive imaging removes the small physical pinhole aperture that otherwise blocks photons and introduces diffraction blur. The detector half-angle ($\theta_{\max}/2$) defines the maximum accessible momentum transfer, $q_{\max}=\frac{4\pi}{\lambda}\sin(\theta_{\max}/2)$. If the full reciprocal-space diameter of the detector is used, the corresponding resolution is $d \simeq 2\pi/q_{\max}$. For a $10~\mathrm{cm}$ detector at $L=3.5~\mathrm{m}$ using a $50~\mathrm{keV}$ probe, $q_{\max} \simeq 3.6~\mathrm{nm}^{-1}$, giving $d\simeq1.7~\mathrm{nm}$.

\textit{Single-shot IDI from HBT correlations---}%
This geometry illustrates the potential of diffraction imaging, but full transverse coherence generally requires an XFEL or a similarly coherent probe. The most accessible x-ray sources are instead laser-driven backlighters. Their picosecond duration can freeze hydrodynamic motion, and their hard x-ray photons can penetrate dense targets, but their spatial incoherence prevents adoption of CDI and requires using an incoherent imaging technique. 

An incoherent incident beam can be modeled as a superposition of independent coherent modes with random phases. Each individual mode scatters from the target to form a distinct coherent diffraction field, as described by Eq.~(\ref{eq:field}). At the detector, the incoherent addition of these overlapping fields washes out macroscopic diffraction fringes, yielding a seemingly random intensity speckle pattern, shown \eg in Fig.~\ref{fig:setup}. However, because every constituent mode diffracts from the identical object, the target's underlying spatial frequencies remain deterministically encoded within the speckle statistics. The object's Fourier modes can therefore be statistically extracted by correlating the spatial intensity fluctuations across different reciprocal-space coordinates.

To formalize this mathematically, the relevant second-order intensity correlation between two detector coordinates $\mathbf q_1$ and $\mathbf q_2$ is
\begin{equation}
    g^{(2)}(\mathbf q_1,\mathbf q_2)
    \equiv
    \frac{
    \left\langle I_s(\mathbf q_1)I_s(\mathbf q_2)\right\rangle
    }{
    \left\langle I_s(\mathbf q_1)\right\rangle
    \left\langle I_s(\mathbf q_2)\right\rangle
    } = 1+ 
    \frac{
    \left|\Gamma_s(\mathbf q_1,\mathbf q_2)\right|^2
    }{
    \left\langle I_s(\mathbf q_1)\right\rangle
    \left\langle I_s(\mathbf q_2)\right\rangle
    }    ,
    \label{eq:g2_definition}
\end{equation}
where $\Gamma_s(\mathbf q_1,\mathbf q_2)$ is the mutual coherence of the scattered field. In this spatial-imaging geometry, the scattered mutual coherence is given by
\begin{align}
    \Gamma_s(\mathbf q_1,\mathbf q_2)
    = &
    \iint J_\mathrm{in}(\mathbf r_1,\mathbf r_2)
    \Delta\phi^*(\mathbf r_1)\Delta\phi(\mathbf r_2)
    \nonumber\\
    &\times
    e^{i\mathbf q_1\cdot\mathbf r_1
    -i\mathbf q_2\cdot\mathbf r_2} d^2\mathbf r_1 d^2\mathbf r_2,
    \label{eq:gamma_general}
\end{align}
where $J_\mathrm{in}(\mathbf r_1,\mathbf r_2)
    =
    \left\langle
    E_{\rm in}^*(\mathbf r_1)E_{\rm in}(\mathbf r_2)
    \right\rangle$
is the incident mutual intensity. 
For a spatially incoherent source, $E_\mathrm{in}(\mathbf r) = \sqrt{I_\mathrm{in}(\mathbf r)}e^{i\chi(\mathbf r)}$, the phases at distinct object-plane points are uncorrelated: $\left\langle e^{i[\chi(\mathbf r_2)-\chi(\mathbf r_1)]} \right\rangle \simeq A_c\delta(\mathbf r_1-\mathbf r_2)$, where $A_c$ is the normalization factor (corresponding to the transverse coherence area) such that $J_\mathrm{in}(\mathbf r,\mathbf r) \to I_\mathrm{in}(\mathbf r)$. Then the scattered mutual coherence reduces to
\begin{equation}
    \Gamma_s(\Delta\mathbf q) = A_c \int I_{\rm in}(\mathbf r) |\Delta\phi(\mathbf r)|^2 e^{-i\Delta\mathbf q\cdot\mathbf r}\,d^2\mathbf r,
    \label{eq:fourier_reduction}
\end{equation}
which is a spatial Fourier transform of $\rho_{\rm IDI}(\mathbf r)=|\Delta\phi(\mathbf r)|^2$ provided that $I_{\rm in}(\mathbf r)$ is uniform. Thus, the random incident phase does not erase the object information; rather, the HBT correlation isolates the squared Fourier modulus of the density perturbation.
If the incident beam possesses partial spatial coherence,  the spatial object is then convolved with the source coherence profile, \eg a Gaussian mutual coherence function, which leads to a decrease of speckle contrast at high spatial frequencies.

The formal HBT covariance requires an ensemble average over many statistically independent detections. The plasma state is highly transient within the detector refresh time, so the ensemble average is replaced by a spatial average over detector-pixel pairs with the same separation $\Delta\mathbf q$. The slowly varying intensity envelope is removed by forming the normalized speckle fluctuation $J(\mathbf q)=I(\mathbf q)/\bar I(\mathbf q)-1$, where $\bar I(\mathbf q)$ is an average local intensity envelope. The single-shot estimator is
\begin{equation}
    {g}^{(2)}(\Delta\mathbf q)-1
    =
    \frac{1}{N_{\Delta q}}
    \sum_{\mathbf q\in\Omega_{\Delta q}}
    J(\mathbf q)J(\mathbf q+\Delta\mathbf q),
    \label{eq:single_shot_g2_estimator}
\end{equation}
where $\Omega_{\Delta q}$ is the set of detector-pixel pairs separated by $\Delta\mathbf q$, and $N_{\Delta q}$ is the total count. 

The spatial-ergodic replacement is accurate only when the detector samples many independent speckles. In reciprocal space, an object of transverse size $D_{\rm obj}$ produces a characteristic speckle width $\Delta q_{\rm speckle}\simeq2\pi/D_{\rm obj}$. For a detector of radius $R_{\det}$ at distance $L$, $q_{\max}=(4\pi/\lambda)\sin(\theta_{max}/2)\simeq2\pi R_{\det}/(\lambda L)$, so the number of independent speckles across the full detector diameter scales as $N_{\rm speckle}\sim(2q_{\max}/\Delta q_{\rm speckle})^2$. Equivalently, the detector-plane speckle width is $s_{\rm speckle}\simeq\lambda L/D_{\rm obj}$. The pixel pitch $p$ must sample each speckle, typically with $n_s = s_{\rm speckle}/p\gtrsim2\text{--}3$, giving $D_{\rm obj}\lesssim\lambda L/[(2\text{--}3)p]$. This sampling condition is restrictive for hard x-ray detectors because decreasing $\lambda$ increases the accessible scattering bandwidth but decreases the speckle size.

The geometric parameters ($R_{\det}, L, \lambda$) set the upper bound $q_{\max}$ and the corresponding full-bandwidth scale $d \simeq \pi/q_{\max}$. However, satisfying the speckle Nyquist condition and securing a large formal speckle count are not sufficient if the target scattering power does not populate those spatial frequencies.
The information-bearing speckle count also depends on the object's spectral morphology. A smooth density profile concentrates its scattering power near $q=0$, leaving the high-$q$ speckles weak. Conversely, a broadband or turbulent density perturbation scatters power widely, providing the extended Fourier spectrum $S(\mathbf{q}) = |\mathcal{F}\{\rho_{\mathrm{IDI}}(\mathbf{r})\}|^2$ needed to overcome the noise floor across the useful angular field of view~\cite{trost2023speckle}.

Ultimately, the effective spatial resolution is bounded not by the detector edge alone, but by the largest reliable correlation separation $\Delta{q}$. For the second-order correlation $g^{(2)}(\Delta\mathbf{q}) \sim \left\langle J(\mathbf{q})J(\mathbf{q}+\Delta\mathbf{q})\right\rangle_{\mathbf{q}}$, the statistical bottleneck is the number of usable intensity pairs. Because the geometric overlap of accessible pixel pairs decreases as $\Delta\mathbf{q}$ approaches the detector bandwidth, high-frequency correlations are less reliable. 
%
The choice of $\Delta q_{max}$ therefore represents a resolution--fidelity tradeoff rather than a purely geometric cutoff. Increasing $\Delta q_{max}$ admits higher spatial frequencies and can sharpen the nominal reconstruction, but the added modes are estimated from fewer statistically independent pixel pairs and generally carry lower scattering power. If these low-fidelity modes are enforced too strongly, they introduce inconsistent Fourier constraints, producing reduced image contrast, ringing, and spurious fine-scale structure. On the other hand, choosing $\Delta q_{max}$ too conservatively improves the reliability of the constraints but smooths the recovered image. The optimal cutoff is thus the largest reciprocal-space radius for which the measured Fourier magnitude and the higher-order phase constraints remain mutually consistent under the positivity and support constraints.

\textit{Phase retrieval using third-order correlation---}%
To reconstruct the real-space object $\rho_{\rm IDI}$, the missing Fourier phase must be retrieved. While CDI relies entirely on iterative projection algorithms, IDI can directly extract deterministic phase constraints using higher-order intensity correlations~\cite{peard2023ab,bojer2024phase}. For chaotic Gaussian radiation, the normalized third-order intensity correlation $    g^{(3)}(\mathbf q_1,\mathbf q_2,\mathbf q_3)\equiv\frac{\left\langle I(\mathbf q_1)I(\mathbf q_2)I(\mathbf q_3)\right\rangle}{\left\langle I(\mathbf q_1)\right\rangle \left\langle I(\mathbf q_2)\right\rangle
\left\langle I(\mathbf q_3)\right\rangle}$ is related to the second order correlation via
\begin{align}
g^{(3)}
&=1
+|g^{(1)}_{12}|^2
+|g^{(1)}_{23}|^2
+|g^{(1)}_{31}|^2
\nonumber\\
&\qquad +
2|g^{(1)}_{12}g^{(1)}_{23}g^{(1)}_{31}|
\cos\Phi_{123},
\label{eq:g3}
\end{align}
where the first-order correlation $g^{(1)}_{ij} = |g^{(1)}_{ij}|\exp(i\varphi_{ij})$ can be obtained using the Siegert relation $g^{(2)} = 1+|g^{(1)}|^2$ for incoherent light. The final term isolates the closure phase
\begin{equation}
\Phi_{123} = \varphi_{12}+\varphi_{23}+\varphi_{31},
\label{eq:closure}
\end{equation}
which corresponds to the phase of the bispectrum $B(\mathbf q_1,\mathbf q_2) = \widetilde{\rho}_{\rm IDI}(\mathbf q_1) \widetilde{\rho}_{\rm IDI}(\mathbf q_2) \widetilde{\rho}_{\rm IDI}^{*}(\mathbf q_1+\mathbf q_2)$.
Thus, we can use the bispectrum to retrieve the Fourier phases using Eq.~(\ref{eq:g3}). 

Note, however, that the experimental observable $g^{(3)}$ can only yield $\cos\Phi_{123}$ and the sign of the local closure phase remains ambiguous. Thus, a recursive phase-unwrapping algorithm or a global optimization should be used to reconstruct the phase map via minimizing the joint least-squares error between the trial object and the $g^{(2)}$ modulus and $g^{(3)}$ cosine constraints.

\begin{figure}[t]
	\centering
	\includegraphics[width=\columnwidth]{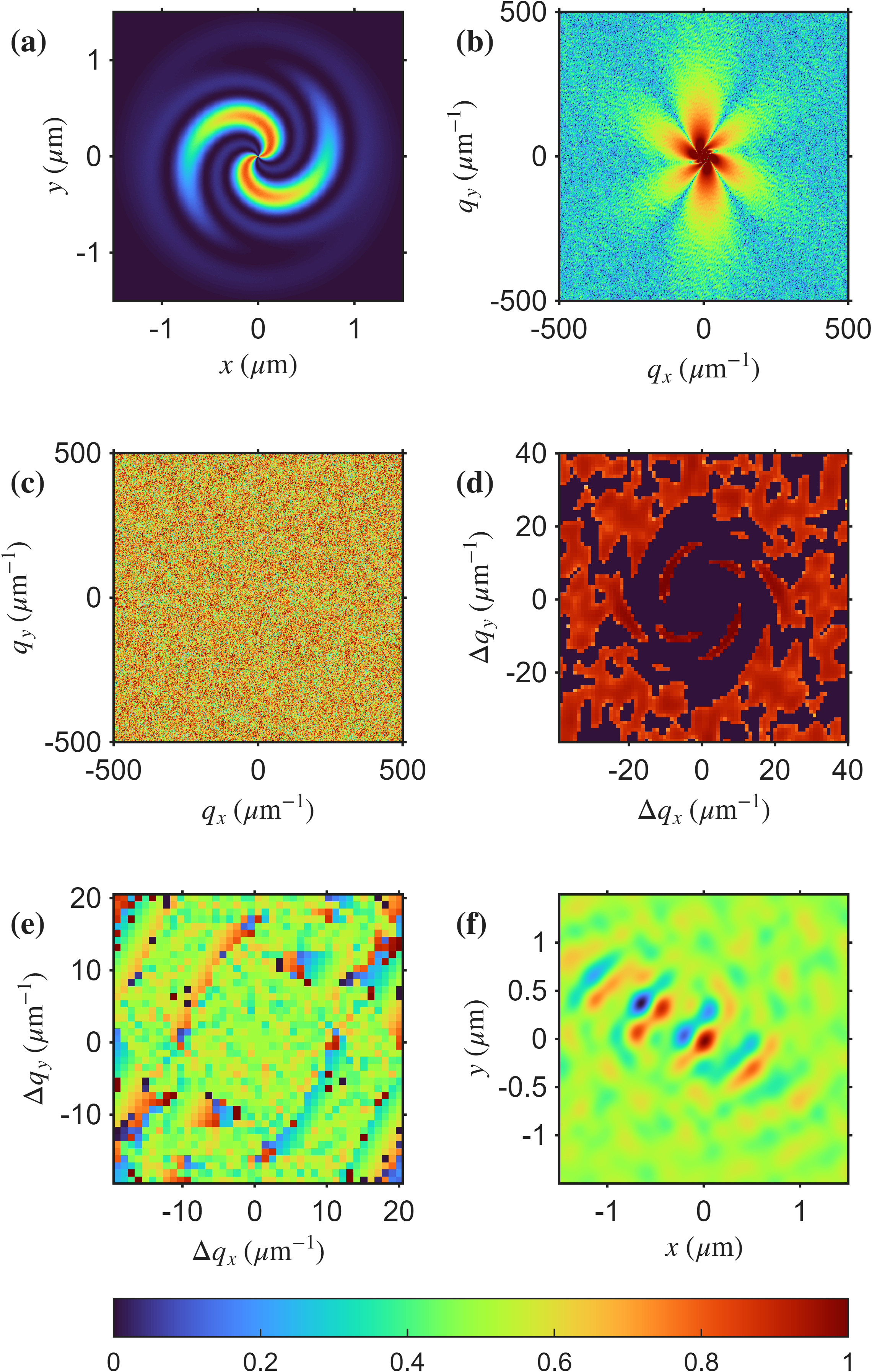} 
	\caption{Single-shot IDI phase retrieval. (a) Kelvin-Helmholtz effective object. (b) Ideal coherent diffraction pattern. (c) Speckled intensity. (d) Fourier magnitude from $g^{(2)}$. (e) Phase recovered via $g^{(3)}$. (f) Final reconstruction.
	}
	\label{fig:numer}
\end{figure}

\textit{Numerical demonstration---}%
To illustrate the single-shot IDI framework, we simulate a $2~\mu\mathrm{m}$-diameter ($D_{\rm obj}$) Kelvin-Helmholtz effective phase object probed by a spatially chaotic $50~\mathrm{keV}$ x-ray pulse ($\lambda = 0.0248~\mathrm{nm}$). 
The object produces a total accumulated phase shift of $0.6~\mathrm{rad}$, with its local value is low enough that refractive ray bending is neglected here. If stronger density gradients introduce appreciable aberrations, the transmitted incident beam could in principle serve as an adaptive-optics reference for wavefront correction~\cite{qu2026polarimetry}, at the cost of a more elaborate reconstruction. The scattered field is recorded at $z=3.5~\mathrm{m}$ on a $10~\mathrm{cm}$-wide detector with $p=13.5~\mu\mathrm{m}$ pixels. The far-field speckle width, $s_{\mathrm{speckle}}\simeq\lambda z/D_{\mathrm{obj}}\simeq43.4~\mu\mathrm{m}$, is oversampled at $s_{\rm speckle}/p\simeq3.2$ pixels, providing approximately $\times10^7$ independent spatial modes to ensure robust statistical averaging.

The object structure is displayed in Fig.~\ref{fig:numer}(a). Under ideal, fully coherent illumination, this object would yield a Fraunhofer diffraction pattern shown in Fig.~\ref{fig:numer}(b), which forms the basis for standard CDI. However, the chaotic illumination prescribed by the IDI framework yields a modulated speckle intensity pattern shown in Fig.~\ref{fig:numer}(c). By itself, this chaotic speckle realization bears no direct resemblance to the target morphology or its ideal diffraction pattern, illustrating why the raw intensity is not an image-bearing observable for incoherent backlighters. 

The structural information is recovered through intensity correlations. We first normalize the raw intensity to obtain a slowly varying envelope, $J(\mathbf{q})=I(\mathbf{q}) /I_{\mathrm{env}}(\mathbf{q})-1$, rendering the spatial statistics approximately stationary. The second-order spatial correlation, extracted via $|g^{(2)}(\Delta\mathbf{q})-1|$, yields the Fourier magnitude of the target. As shown in Fig.~\ref{fig:numer}(d), the low-$q$ modes of this correlation recover the signature of the ideal diffraction pattern. To retrieve the missing spatial phase [Fig.~\ref{fig:numer}(e)], a finite high-signal domain ($\Delta q_{max} = 18~\mu\mathrm{m}^{-1}$) is selected to construct spatial triplets for the $g^{(3)}$ bispectrum. A global multi-start optimizer then enforces these $g^{(3)}$ closure phases alongside standard positivity and support constraints. 

The final IDI reconstruction is shown in Fig.~\ref{fig:numer}(f). It captures the main spiral morphology and the $\unit[\sim0.1]{\mu m}$ arm width of the density perturbation, though the finite correlation bandwidth smooths the sharpest interfaces. The selected cutoff $\Delta q_{max}$ balances spatial resolution ($d\sim\pi/\Delta q_{max}\simeq 90~\mathrm{nm}$) against image contrast. While incorporating a larger $\Delta q_{max}$ could theoretically enhance resolution, the corresponding drop in the $g^{(3)}$ signal-to-noise ratio introduces unreliable Fourier components that degrade the final contrast. Ultimately, this simulation demonstrates that complex morphologies can be retrieved in a single shot, with the achieved resolution fundamentally dictated by the measurable correlation bandwidth rather than geometric detector edges.

\textit{3D reconstruction---}
Each IDI view reconstructs a two-dimensional projected SAXS object, $\rho_{\rm IDI,\theta}=I_{b,\theta}|\Delta\phi_\theta|^2$. Mapping the full three-dimensional electron-density perturbation $\delta n_e(\mathbf r)$ relies on the standard Fourier-slice theorem, where the two-dimensional Fourier transform of each projection samples a central plane of the 3D target spectrum. Thus, multi-view IDI can constrain 3D structure only through these projected correlation objects. 

Because microscopic plasma fluctuations are not reproducible from shot to shot, any 3D extension requires simultaneous multi-beam x-ray probing. Such a sparse multi-angle implementations transitions the inverse problem into a joint constrained optimization over the multi-view IDI morphology, the measured single-shot covariances, and real-space support or symmetry constraints. Given this sparse tomography, the most robust near-term 3D observables will not be arbitrary voxel-resolved turbulence, but rather targeted statistical metrics such as anisotropic power spectra and dominant instability modes.

\textit{Discussion---}%
Translating the single-shot IDI framework to an ICF plasma object requires navigating several physical constraints. 
First, the diagnostic could be constrained by the correlated photon budget. The signal-to-noise ratio (SNR) of the detection scales with the mean number of detected signal photons per speckle, $\bar{n}$, \ie $\mathrm{SNR} \propto \bar{n} \sqrt{N_{\mathrm{pair}}^{\mathrm{eff}}}$. A large detector pixel number combined with $N_{\mathrm{pair}}^{\mathrm{eff}} \sim 10^3$--$10^5$ independent pairs could allow viable reconstruction using laser-driven backlighters. However, any uncorrelated plasma self-emission yielding $\bar{n}_{\mathrm{bg}}$ photons per speckle dilutes the $g^{(2)}$ and $g^{(3)}$ correlation contrast by a factor of $[\bar{n}/(\bar{n} + \bar{n}_{\mathrm{bg}})]^{2,3}$. This dictates that IDI is best deployed during low-emission pre- or post-stagnation frames, utilizing high-Z spectral filtering to isolate the hard x-ray probe from softer thermal backgrounds. 

Second, laser-driven backlighters are not monochromatic. A finite spectral bandwidth $\Delta\lambda$ causes far-field speckles to smear radially, degrading the correlation contrast at the detector edge by an amount $\Delta R \simeq R_{\mathrm{det}}(\Delta\lambda/\lambda)$. For the high-frequency speckle contrast to survive, this chromatic blur must not exceed the speckle width $s_{\mathrm{speckle}}$. This imposes a condition on the maximum allowable spectral bandwidth:
\begin{equation}
    \frac{\Delta\lambda}{\lambda} \lesssim \frac{s_{\mathrm{speckle}}}{R_{\mathrm{det}}} \simeq \frac{d}{D_{\mathrm{obj}}}.
\end{equation}
To achieve a resolution of $d = 0.1~\mathrm{\mu m}$ across a $D_{\mathrm{obj}} = 2~\mu\mathrm{m}$ target, the required probe bandwidth is $\Delta\lambda/\lambda \lesssim 5\%$. This tolerance is consistent with the natural linewidth of standard $K\alpha$ emission, allowing the use of existing sources without dispersive filtering. Nevertheless, future integration of narrower-bandwidth inverse-Compton or betatron sources would further suppress chromatic blur and extend the accessible high-resolution bandwidth.

Third, spatial resolution is subject to backlight chromaticity and hydrodynamic temporal smearing. Hydrodynamic evolution dictates an upper bound on the exposure time: if the density structures move at a velocity $v$, resolving a feature size $d$ requires $v\tau < d$. For a turbulent plasma interface moving at $100~\mathrm{km/s}$ ($0.1~\mathrm{\mu m/ps}$), capturing a $0.1~\mathrm{\mu m}$ feature requires an x-ray pulse shorter than $1~\mathrm{ps}$. A finite spectral bandwidth $\Delta\lambda$ also smears the diffraction speckles because multiple modes produce different speckle patterns by possibly multiple orders of magnitudes. While it can be mitigated by multi-megapixel detectors to provide the statistical averaging, this framework strongly motivates the continued development of ultra-short hard x-ray backlighters to completely bypass the temporal-mode contrast penalty.

	\begin{acknowledgments}
		KQ and NJF were supported by NNSA Grant No. DE-NA0004167,  NSF Grant No. PHY- 2308829, and the  Center for Magnetic Acceleration, Compression, and Heating (MACH), part of the U.S. DOE-NNSA Stewardship Science Academic Alliances Program under Cooperative Agreement No. DE-NA0004148. DB was supported by the JSPS Bilateral Program Number JPJSBP120257718.
	\end{acknowledgments}  

\bibliography{references}

\end{document}